\documentclass[aps,prd,preprint,superscriptaddress,amsmath,amssymb,showpacs]{revtex4-1}
\usepackage{dcolumn}
\usepackage{graphicx}
\usepackage{float}
\usepackage{physics}
\usepackage[colorlinks=true,allcolors=blue]{hyperref}

\begin{document}

\title{Imaginary potential and thermal width in the spinning black hole background from holography}

\author{Zhou-Run Zhu }
\email{zhuzhourun@zknu.edu.cn}
\affiliation{School of Physics and Telecommunications Engineering, Zhoukou Normal University, Zhoukou 466001, China}

\author{Sheng Wang}
\email{shengwang@mails.ccnu.edu.cn}
\affiliation{Institute of Particle Physics and Key Laboratory of Quark and Lepton Physics (MOS), Central China Normal University, Wuhan 430079, China}

\author{Yang-Kang Liu}
\email{ykl@mails.ccnu.edu.cn}
\affiliation{Institute of Particle Physics and Key Laboratory of Quark and Lepton Physics (MOS), Central China Normal University, Wuhan 430079, China}

\author{Defu Hou}
\email{houdf@mail.ccnu.edu.cn}
\affiliation{Institute of Particle Physics and Key Laboratory of Quark and Lepton Physics (MOS), Central China Normal University, Wuhan 430079, China}

\begin{abstract}
In this study, we investigate the imaginary potential of heavy quarkonium in the spinning black hole background. Then we estimate the thermal width, which is determined by the imaginary part of the finite temperature potential. In the ultra-local description, the boosted fluid represents a globally rotating fluid. Using a holographic approach, we systematically analyze how boost parameter influences these quantities. Our results reveal that increasing boost parameter causes the imaginary potential to emerge at smaller interquark distances, suggesting that boost parameter accelerates quarkonium melting. Furthermore, we find that boost parameter enhances the thermal width, indicating greater instability of the bound state at higher boost parameter. Notably, we observe that the effect of boost parameter on quarkonium dissociation is more pronounced when the axis of the quark-antiquark pair is transverse to the direction of boost parameter.

\end{abstract}
\maketitle

\section{Introduction}\label{sec:01_intro}

Heavy quarkonium consists of a heavy quark and a heavy antiquark, serving as a powerful probe for understanding the properties of quark-gluon plasma (QGP). The melting of heavy quarkonium is one of the primary experimental indications of the formation of QGP. One direct evidence of this is the suppression of heavy quarkonium due to the color screening \cite{Matsui:1986dk}. Another reason for the suppression is the imaginary part of the potential. Some related researches \cite{Laine:2006ns,Beraudo:2007ky,Burnier:2009yu} have suggested that the imaginary potential may play a more critical role in explaining thermal properties of quarkonium within the strongly coupled medium. The imaginary potential can explain the dissociation and decay of the bound state. Additionally, it can be used to estimate the thermal width of quarkonium related to thermal decay \cite{Brambilla:2010vq,Brambilla:2008cx,Margotta:2011ta}.

The non-central heavy ion collisions experiments have shown the generation of nonzero angular momentum \cite{Liang:2004ph,Becattini:2007sr,Baznat:2013zx,STAR:2017ckg,Jiang:2016woz}. While part of the angular momentum is carried away by spectator partons, a significant portion is transferred to the strongly coupled plasma \cite{Baznat:2015eca,Kharzeev:2015znc}. In \cite{STAR:2017ckg}, the results from Star Collaboration on the $\Lambda$ and $\Lambda^-$ baryons provide strong evidence and yield an angular velocity average value $\omega \sim$ 6 MeV. The prediction of heavy-ion collisions hydrodynamic simulations \cite{Jiang:2016woz} suggests that the magnitudes of angular velocity can be larger, $\omega \sim$ (20-40) MeV. These nonzero angular velocity values result in relativistic rotation of the strongly coupled plasma. Investigating the influence of rotation on the thermodynamics of heavy quarkonium is an attractive research topic.

AdS/CFT correspondence \cite{Maldacena:1997re,Witten:1998qj,Gubser:1998bc} serves as a powerful non-perturbative method for investigating the properties of strongly coupled systems and offers valuable insights into the behavior of heavy quarks in strongly coupled plasma. In the context of holography, quarks correspond to the endpoints of open strings that are embedded in the bulk of AdS spacetime. A quark-antiquark pair is represented by a string that connects the two endpoints. This string forms on a U-shaped configuration, and its dynamics are described by the Nambu-Goto action. By using AdS/CFT correspondence, Noronha and Dumitru first calculate the imaginary potential in $\mathcal{N} = 4$ super-Yang Mills theory from thermal fluctuation \cite{Noronha:2009da}. The study of the imaginary potential of heavy quarkonium has gained significant attention. The authors of Ref.\cite{Finazzo:2013rqy} discuss the imaginary potential, and calculate the thermal width from imaginary potential in the static background. The dissociation of heavy quarkonium in the moving case has been studied in \cite{Liu:2006nn}. The usual is to boost the particle pair into a reference frame where they appear at rest, while the hot wind moves toward them. Additionally, the imaginary potential in the moving cases has been studied in \cite{Finazzo:2014rca,Ali-Akbari:2014vpa}. In Refs. \cite{Fadafan:2013coa,BitaghsirFadafan:2015yng}, the authors investigate the influence of finite 't Hooft coupling corrections on the imaginary potential and the thermal width. The imaginary potential and the thermal width within AdS/QCD can be seen in \cite{Sadeghi:2014zya,Braga:2016oem,Zhang:2019qhm}. The imaginary potential in the rotating matter has been studied in \cite{Zhang:2023psy}. In Ref. \cite{BitaghsirFadafan:2013vrf}, the authors explore the imaginary potential in the anisotropic plasma. Other related works can be seen in \cite{Zhang:2016tem,Zhang:2018fpe,Zhao:2019tjq,Kioumarsipour:2021zyg,Kioumarsipour:2024spj,Albacete:2008dz}.

In this work, we want to explore the imaginary potential and estimate the thermal width from imaginary potential in the spinning black hole background from holography. It should be mentioned that the authors of Ref. \cite{Zhang:2023psy} extend the static black hole to a local rotating by the Lorentz transformation \cite{Erices:2017izj,BravoGaete:2017dso,Awad:2002cz}, and calculate the imaginary potential in the rotating background. From this Lorentz transformation, the resulting black hole is a (hyper) cylindrical surface about the symmetry axis, and just represents a small area around the rotating radius $L$ with a domain range less than 2$\pi$. Additionally, the string breaking and running coupling behavior of heavy quark-antiquark pairs are systematically investigated in a rotating background \cite{Zhou:2023qtr}. Furthermore, Refs. \cite{Arefeva:2020jvo,Golubtsova:2021agl,Arefeva:2020knc} explore heavy quark dynamics in rotating quark-gluon plasma, with particular focus on the drag force and jet quenching parameter in the rotational background.

It is meaningful to examine the imaginary potential and estimate the thermal width from imaginary potential in the spinning black hole background from holography \cite{Hawking:1998kw,Gibbons:2004ai,Gibbons:2004js,Garbiso:2020puw,Amano:2023bhg}. As discussed in \cite{Garbiso:2020puw}, Myers-Perry black holes (spinning black holes) are the solutions in higher-dimensional spacetimes under Einstein gravity, which describe a rotating black hole with independent angular momentum. The boundary of the spacetime is typically compact, which leads to a gauge theory that lives on a spacetime $S^3 \times \mathbb{R}$. In the limit of large black hole, the Myers-Perry line element becomes a Schwarzschild black brane boosted along $x_3$ direction. The component associated with the black hole dominates and exhibits correspondence with a rigidly rotating fluid in $S^3 \times \mathbb{R}$. When projecting this flow onto flat spacetime, one observes a fluid exhibiting markedly non-trivial vorticity and expansion. Local flow analysis reveals that at leading order (LO) in the large black hole expansion, the fluid manifests as a uniformly boosted fluid, while the next-to-leading order (NLO) correction incorporates a rotation about the boost direction \cite{Amano:2023bhg}. When investigating rotational effects, we consider zooming in on a small patch and will observe a uniformly streaming fluid. In the ultra-local description, the boosted fluid represents a globally rotating fluid. In the dual field theory, the planar black brane is particularly relevant for understanding the behavior of the QGP. Inspired by this, we aim to examine the impact of boost parameter on the imaginary potential and estimate the thermal width from imaginary potential in the Myers-Perry black hole background.

The paper is organized as follows. In Sec.~\ref{sec:02}, we briefly review the spinning black hole background. In Sec.~\ref{sec:03}, we explore the imaginary potential and estimate the thermal width from imaginary potential in the Myers-Perry black hole background. In Sec.~\ref{sec:04}, we give the conclusion and discussion.

\section{Spinning Myers-Perry black hole background}\label{sec:02}

In this work, we aim to investigate the the imaginary potential and estimate the thermal width from imaginary potential in the Myers-Perry black hole background. We first briefly review the metric of five-dimensional spinning black hole background, which is proposed by Hawking et al \cite{Hawking:1998kw}

\begin{equation}
\label{eqc1}
\begin{split}
ds^{2} & =-\frac{\Delta}{\rho^{2}}(dt_{H}-\frac{a\sin^{2}\theta_{H}}{\Xi_{a}}d\phi_{H}-\frac{b\cos^{2}\theta_{H}}{\Xi_{b}}d\psi_{H})^{2}\\
 & +\frac{\Delta_{\theta_{H}}\sin^{2}\theta_{H}}{\rho^{2}}(adt_{H}-\frac{r_{H}^{2}+a^{2}}{\Xi_{a}}d\phi_{H})^{2}+\frac{\Delta_{\theta_{H}}\cos^{2}\theta_{H}}{\rho^{2}}(bdt_{H}-\frac{r_{H}^{2}+b^{2}}{\Xi_{b}}d\psi_{H})^{2}+\frac{\rho^{2}}{\Delta}dr_{H}^{2}\\
 & -\frac{\rho^{2}}{\Delta_{\theta_{H}}}d\theta_{H}^{2}+\frac{1+\frac{r_{H}^{2}}{L^{2}}}{r_{H}^{2}\rho^{2}}(abdt_{H}-\frac{b\left(r^{2}+a^{2}\right)\sin^{2}\theta_{H}}{\Xi_{a}}d\phi_{H}-\frac{a\left(r^{2}+b^{2}\right)\cos^{2}\theta_{H}}{\Xi_{b}}d\psi_{H})^{2},
 \end{split}
\end{equation}
with
 \begin{equation}
\label{eqc11}
\begin{split}
\Delta=\frac{1}{r_{H}^{2}}(r_{H}^{2}+a^{2})(r_{H}^{2}+b^{2})(1+\frac{r_{H}^{2}}{L^{2}})-2M,\\
 \Delta_{\theta_{H}}=1-\frac{a^{2}}{L^{2}}\cos^{2}\theta_{H}-\frac{b^{2}}{L^{2}}\sin^{2}\theta_{H},\\
 \rho=r_{H}^{2}+a^{2}\cos^{2}\theta_{H}+b^{2}\sin^{2}\theta_{H},\\
 \Xi_{a}=1-\frac{a^{2}}{L^{2}},\\
\Xi_{b}=1-\frac{b^{2}}{L^{2}},\\
 \end{split}
\end{equation}
where $a$ and $b$ are the angular momentum parameters. In this work, we consider spinning Myers-Perry black hole case, namely $a=b$ case \cite{Gibbons:2004ai,Gibbons:2004js}. In addition, $\phi_H$, $\psi_H$ and $\theta_H$ represent angular Hopf coordinates. $t_H$ denotes time, $L$ is the AdS radius, and $r_H$ represents AdS radial coordinate. $M$ is the mass.

After adopting the more convenient coordinates \cite{Murata:2008xr}
 \begin{equation}
\label{eqc111}
\begin{split}
t=t_{H},\\
 r^{2}=\frac{a^{2}+r_{H}^{2}}{1-\frac{a^{2}}{L^{2}}},\\
\theta=2\theta_{H},\\
 \phi=\phi_{H}-\psi_{H},\\
\psi=-\frac{2at_{H}}{L^{2}}+\phi_{H}+\psi_{H},\\
b=a,\\
\mu=\frac{M}{(L^{2}-a^{2})^{3}},\\
 \end{split}
\end{equation}
one can rewrite the metric (\ref{eqc1}) as
\begin{equation}
\label{eqc2}
\begin{split}
ds^{2}=-(1+\frac{r^{2}}{L^{2}})dt^{2}+\frac{dt^{2}}{G(\text{r)}}+\frac{r^{2}}{4}((\sigma^{1})^{2}+(\sigma^{2})^{2}+(\sigma^{3})^{2})+\frac{2\mu}{r^{2}}(dt+\frac{a}{2}\sigma^{3})^{2},
 \end{split}
\end{equation}
with
 \begin{equation}
\label{eqc3}
\begin{split}
G(r)=1+\frac{r^{2}}{L^{2}}-\frac{2\mu(1-\frac{a^{2}}{L^{2}})}{r^{2}}+\frac{2\mu a^{2}}{r^{4}},\\
\mu=\frac{r_{h}^{4}(L^{2}+r_{h}^{2})}{2L^{2}r_{h}^{2}-2a^{2}(L^{2}+r_{h}^{2})},\\
\sigma^{1}=-sin\psi dtd\theta+cos\psi sin\theta d\phi,\\
\sigma^{2}=cos\psi d\theta+sin\psi sin\theta d\phi,\\
\sigma^{3}=d\psi+cos\theta d\phi,\\
 \end{split}
\end{equation}
where
\begin{equation}
\label{eqc4}
\begin{split}
-\infty<t<\infty,\ r_{h}<r<\infty,\ 0\leq\theta\leq\pi,\ 0\leq\phi\leq2\pi,\ 0\leq\psi\leq4\pi.
 \end{split}
\end{equation}

The planar black brane can be obtained by the coordinate transformation \cite{Garbiso:2020puw}
 \begin{equation}
\label{eqc5}
\begin{split}
t=\tau,\\
\frac{L}{2}(\phi-\pi)=x_{1},\\
\frac{L}{2}tan(\theta-\frac{\pi}{2})=x_{2},\\
\frac{L}{2}(\psi-2\pi)=x_{3},\\
r=\tilde{r},\\
 \end{split}
\end{equation}
where $(\tau, \widetilde{r}, x_1, x_2, x_3)$ denote new coordinates. These coordinates can be scaled with a factor $\beta$
 \begin{equation}
\label{eqc6}
\begin{split}
\tau\rightarrow\beta^{-1}\tau,\\
x_{1}\rightarrow\beta^{-1}x_{1},\\
x_{2}\rightarrow\beta^{-1}x_{2},\\
x_{3}\rightarrow\beta^{-1}x_{3},\\
\tilde{r}\rightarrow\beta\tilde{r},\\
\tilde{r_{h}}\rightarrow\beta\tilde{r_{h}}.(\beta\rightarrow \infty)\\
 \end{split}
\end{equation}

One can derive the metric of a Schwarzschild black brane, which is boosted in the $\tau - x_3$ plane \cite{Garbiso:2020puw}
\begin{equation}
\label{eqc7}
\begin{split}
ds^{2}=\frac{r^{2}}{L^{2}}(-d\tau^{2}+dx_{1}^{2}+dx_{2}^{2}+dx_{3}^{2}+\frac{r_{h}^{4}}{r^{4}(1-\frac{a^{2}}{L^{2}})}(d\tau+\frac{a}{L}dx_{3})^{2})+\frac{L^{2}r^{2}}{r^{4}-r_{h}^{4}}dr^{2},
 \end{split}
\end{equation}
where the Schwarzschild black brane can be recovered when boost parameter vanishes ($a=0$).

Then, we take $z$ ($r=1/z$) as the holographic fifth coordinate for simplified calculations
\begin{equation}
\label{eqc8}
\begin{split}
ds^{2}=\frac{1}{z^{2}L^{2}}[-d\tau^{2}+dx_{1}^{2}+dx_{2}^{2}+dx_{3}^{2}+\frac{z^{4}}{z_{h}^{4}(1-\frac{a^{2}}{L^{2}})}(d\tau+\frac{a}{L}dx_{3})^{2}]+\frac{L^{2}z_{h}^{4}}{z^{2}(z_{h}^{4}-z^{4})}dz^{2}.
 \end{split}
\end{equation}

The expression of temperature is \cite{Garbiso:2020puw}
\begin{equation}
\label{eqb1}
T_h=\frac{\sqrt{L^{2}-a^{2}}}{z_{h}\pi L^{3}},
\end{equation}
where $z_h$ denote the horizon. In the limitation of the large black hole, the angular velocity $\Omega = a/L^2$ \cite{Garbiso:2020puw}. In the following calculations, we take AdS radius $L=1$ for simplicity.

\section{Imaginary potential and thermal width in spinning Myers-Perry black hole background}\label{sec:03}

In this section, we explore the imaginary potential and estimate the thermal width from imaginary potential in the spinning background, which could be calculated from the on-shell action of world sheet. The formulas can be derived by following the steps in Ref. \cite{Noronha:2009da,Finazzo:2013rqy,BitaghsirFadafan:2013vrf}.

The on-shell Nambu-Goto action is
\begin{equation}
\label{eq10}
S= -\frac{1}{2\pi\alpha'} \int d\tau d\sigma \sqrt{-det g_{\alpha \beta}},
\end{equation}
where $g_{\alpha \beta}$ is the determinant of induced metric. $\tau$ and $\sigma$ represents the world sheet coordinates.

In the following discussion, we will explore the impact of boost parameter in different directions. The gravitational metric can be viewed as boosted in the $\tau- x_3$ plane, indicating that the direction of rotation is along the $x_3$ direction. Meanwhile, the axis of the $Q\overline{Q}$ pair in the $x_1 - x_2$ plane is considered a transverse case, whereas the axis of the $Q\overline{Q}$ pair aligned with the $x_3$ is regarded as a parallel case.

In transverse case, the world sheet coordinates can be parameterized by
\begin{equation}
\label{eq11}
\tau=\xi,\ x_{1}=\eta,\ x_{2}=0,\ x_{3}=0,\ z=z(\eta),
\end{equation}

The Lagrangian density can be obtained as
\begin{equation}
\label{eq12}
\mathcal{L}=\sqrt{A(z)+B(z)\dot{z}^{2}},
\end{equation}
with
 \begin{equation}
\label{eq13}
\begin{split}
&A(z)=A(z_\perp)=\frac{1}{z^{4}}-\frac{1}{z_{h}^{4}(1-a^{2})},\\
&B(z)=B(z_\perp)=\frac{z_{h}^{4}}{z^{4}(z_{h}^{4}-z^{4})}-\frac{1}{(z_{h}^{4}-z^{4})(1-a^{2})}.
 \end{split}
\end{equation}

In parallel case, the world sheet coordinates can be parameterized by
\begin{equation}
\label{eq111}
\tau=\xi,\ x_{1}=0,\ x_{2}=0,\ x_{3}=\eta,\ z=z(\eta).
\end{equation}

The expressions of $A(z)$ and $B(z)$ are
 \begin{equation}
\label{eq13}
\begin{split}
&A(z)=A(z_\parallel)=\frac{z_{h}^{4}-z^{4}}{z^{4}z_{h}^{4}},\\
&B(z)=B(z_\parallel)=\frac{z_{h}^{4}}{z^{4}(z_{h}^{4}-z^{4})}-\frac{1}{(z_{h}^{4}-z^{4})(1-a^{2})}.
 \end{split}
\end{equation}

The interquark distance $L$ of $Q\overline{Q}$ is
 \begin{equation}
\label{eq14}
\begin{split}
L=2\int_{0}^{z_{\ast}}\sqrt{\frac{A(z_{\ast})B(z)}{A(z)^{2}-A(z)A(z_{\ast})}}dz,
 \end{split}
\end{equation}
where $z_\ast$ ($0<z_{\ast}<z_h$) is the deepest point of the U-shaped string.

The imaginary potential can be calculated from the thermal fluctuations of the string worldsheet \cite{Noronha:2009da}, One can incorporate thermal fluctuations $\delta z(x)$ about the classical worldsheet configuration $z_c(x)$, where $\delta z(x)$ characterizes thermal excitations of the string
\begin{equation}
    z(x)=z_c(x)\rightarrow z(x)=z_c(x)+\delta z(x).
\end{equation}

In the long wavelength fluctuations approximation for string fluctuations, where the string endpoint $z_{\ast}$ is sufficiently close to the horizon $z_h$, certain fluctuation modes can penetrate the horizon region. The partition function, including thermal fluctuation effects, then takes the form:
\begin{equation}
Z_{str}\sim\int\mathcal{D}\delta z(x)e^{iS_{NG}(z_c(x)+\delta z(x))}.
\end{equation}

We evaluate the $\delta z(x)$ integral numerically by introducing a uniform discretization of the interval $[-\frac{L}{2},\frac{L}{2}]$ with 2N parts, establishing grid points at $x_j=j\Delta{x}=j\frac{L}{2N}$:
\begin{equation}
    Z_{str}\sim \lim_{N\rightarrow \infty}d[\delta z(x_{-N})]\cdots d[\delta z(x_N)]\exp[\frac{i\mathcal{T}\Delta x}{2\pi\alpha^{\prime}}\sum_j\sqrt{A(z_j)+B(z_j)z^{\prime2}_j}],
    \label{z str}
\end{equation}
where $z_j\equiv z(x_j)$, $z_j^{\prime}\equiv\frac{dz(x)}{dx}|_{x=x_j}$. Since the thermal fluctuations are predominantly localized near the bottom of the string (corresponding to $x=0$ where $z=z_{\ast}$), we can expand $z_c(x_j)$ around $x=0$ while retaining only terms up to second order in $x_j$. This approximation is justified by the fact that $z_c^{\prime}(0)=0$ at this point
\begin{equation}
    z_c(x_j)\simeq z_{\ast}+\frac{x_j^2}{2}z_c^{\prime\prime}(0).
    \label{z expansion}
\end{equation}

One can expand $A(z_j)$ and keep  terms up to second order in $x_j^m\delta{z_n}$
\begin{equation}
    A(z_j)\approx A_{\ast}+A_{\ast}^{\prime}\frac{x_j^2}{2}z_c^{\prime\prime}(0)+A^{\prime}_{\ast}\delta z+A^{\prime\prime}_{\ast}\frac{\delta z^2}{2},
    \label{A expansion}
\end{equation}
where $A_{\ast}\equiv A(z_{\ast}),A(z_{\ast})\equiv\frac{\partial A(z)}{\partial z}|_{z=z_{\ast}}$, etc. The exponent in Eq. (\ref{z str}) can be approximated as
\begin{equation}
    S_j^{NG}=\frac{\mathcal{T}\Delta x}{2\pi\alpha^{\prime}}\sqrt{C_1x_j^2+C_2},
    \label{j action}
\end{equation}
where
\begin{align}
    C_1&=\frac{z_c^{\prime\prime(0)}}{2}[2B_{\ast}z_c^{\prime\prime}(0)+A_{\ast}^{\prime}],\nonumber\\
    C_2&=A_{\ast}+\delta zA_{\ast}^{\prime}+\frac{\delta z^2}{2}A_{\ast}^{\prime\prime}.
\end{align}

The analysis reveals that if $S_j^{NG}$ induces an imaginary component in the potential, the argument of the square root in Eq.(\ref{j action}) becomes negative. For the jth contribution, this yields:
\begin{equation}
    I_j\equiv\int_{\delta z_{jmin}}^{\delta z_{jmax}}\exp[\frac{i\mathcal{T}\Delta x}{2\pi\alpha^{\prime}}\sqrt{C_1x_j^2+C_2}],
    \label{j partition}
\end{equation}
where $\delta z_{jmin}$ and $\delta z_{jmax}$ denote the roots of $C_1x_j^2+C_2$ in $\delta{z}$. The integral can be evaluated via saddle-point approximation, with the stationary point occurring when the argument of the square root in Eq. (\ref{j partition}) satisfies the extremum condition.
\begin{equation}
    D(z_j)\equiv C_1x_j^2+C_2(\delta z_j),
\end{equation}
and assumes an extremal value and happens for
\begin{equation}
\delta z=-\frac{A_{\ast}^{\prime}}{A_{\ast}^{\prime\prime}}.
\end{equation}

The condition for the square root to develop an imaginary part requires $x_j<|x_c|$
\begin{equation}
    x_c=\sqrt{\frac{1}{C_1}(\frac{A_{\ast}^{\prime}}{2A_{\ast}^{\prime\prime}}-A_{\ast})}.
\end{equation}

The complete contribution arises from summing individual terms, yielding
$\Pi_jI_j$. This gives the result:
\begin{equation}
    ImV=-\frac{1}{2\pi\alpha^{\prime}}\int_{|x|<x_c}dx\sqrt{-C_1x^2-A_{\ast}+\frac{A^2_{\ast}}{2A_{\ast}^{\prime2}}}.
    \label{x<xc}
\end{equation}

Integration of Eq. (\ref{x<xc}) yields the expression for the imaginary potential $ImV$
 \begin{equation}
\label{eq17}
\begin{split}
ImV= -\frac{\sqrt{\lambda}}{2\sqrt{2}}\sqrt{B(z_{\ast})} (\frac{A'(z_{\ast})}{2A''(z_{\ast})}-\frac{A(z_{\ast})}{A'(z_{\ast})}),
 \end{split}
\end{equation}
where $A'(z_{\ast})$ and $A''(z_{\ast})$ are the first and second derivative of $A(z_{\ast})$, respectively.

To obtain the thermal width, one could derive the expectation value of the imaginary potential using the non-relativistic approximation \cite{Noronha:2009da,Finazzo:2013rqy}
\begin{equation}
    \Gamma = -\langle \psi | Im V | \psi \rangle,
\end{equation}
where
\begin{equation}
\langle \vec{r} | \psi \rangle = \frac{1}{\sqrt{\pi} a_0^{3/2}} e^{-r/a_0},
\end{equation}
is the ground-state wave function of a particle and the particle has a Coulomb-like potential. Thus one can obtain the thermal width $\Gamma$ defined by the imaginary potential.

The thermal width of $Q\overline{Q}$ is
 \begin{equation}
\label{eq18}
\begin{split}
\Gamma/T= -\frac{4}{(a_0 T)^3}\int d\omega \omega^2 e^{\frac{-2\omega}{a_0 T}}\frac{ImV(\omega)}{T},
 \end{split}
\end{equation}
where $\omega=LT$ and $a_0$ is the Bohr radius. There are exact approach and approximate approach to calculate the thermal width \cite{Finazzo:2013rqy,BitaghsirFadafan:2013vrf}. It should be mentioned that the authors of \cite{Zhao:2019tjq} study the effect of gluon condensate on the thermal width by exact approach and approximate approach. The qualitative results from the two approaches are consistent. In this work, we adopt the approximate approach to discuss the effect of boost parameter on the thermal width.

\begin{figure*}[htbp]
    \centering
      \setlength{\abovecaptionskip}{0.1cm}
    \includegraphics[width=0.8\textwidth]{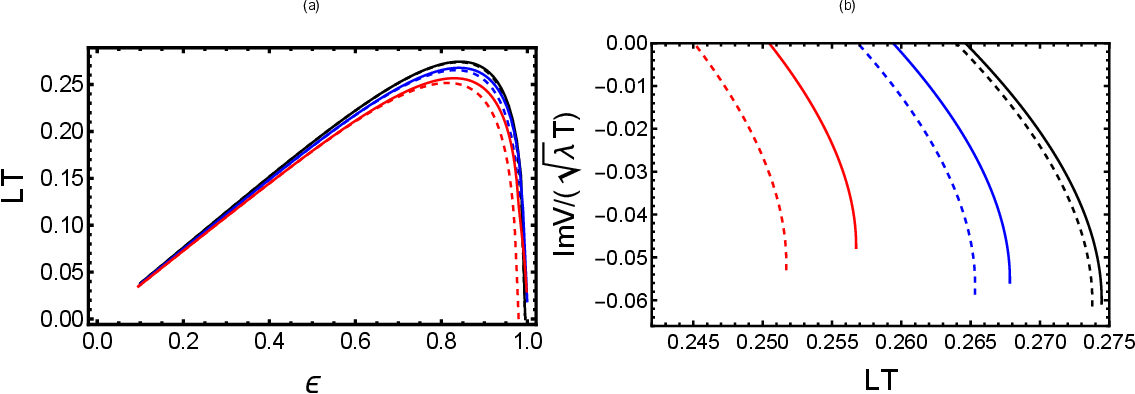}
    \caption{  \label{fig1}  (a) $LT$ versus $\varepsilon$ with different boost parameter $a$. (b) Imaginary potential $ImV/(\sqrt{\lambda}T)$ of quarkonium versus $LT$ with different boost parameter $a$. The black, blue and red line denote $a = 0.1,\ 0.2,\ 0.3$, respectively. The solid line (dashed line) represents the parallel (transverse) case. }
\end{figure*}

We will examine the imaginary potential and estimate the thermal width from imaginary potential in the spinning Myers-Perry black hole from Eqs.(\ref{eq14}), (\ref{eq17}), and (\ref{eq18}). For our calculations, we take $\lambda=1$. According to Ref. \cite{Garbiso:2020puw}, the background geometry features a planar black brane at high temperatures. Therefore, we choose a high temperature $T=100/\pi$. Additionally, the background remains stable when $a < 0.75L$. Thus, we take $a=0.1, 0.2, 0.3$ in the numerical calculations. It should be mentioned that the spinning background is dual to the rotating QGP system. The behavior of QGP flow can be visualized as a uniform motion, characterized by uniformly rotating circles of a Hopf fibration within the spatial component of four-dimensional spacetime \cite{Amano:2023bhg}.

Then we discuss three restrictions on the imaginary potential. Firstly, the term $B(z_\ast)$ in Eq.(\ref{eq17}) must be positive. Secondly, the imaginary potential should be negative. This requirement implies that $(\frac{A'(z_{\ast})}{2A''(z_{\ast})}-\frac{A(z_{\ast})}{A'(z_{\ast})})>0$ in Eq.(\ref{eq17}), leading to the condition $\varepsilon>\varepsilon_{min}$ ($\varepsilon \equiv z_{\ast} /z_h$). The last restriction is related to the the maximum value of $LT$, denoted as $LT_{max}$. In the left panel of Fig.~\ref{fig1}, namely Fig.~\ref{fig1}(a), we draw the interquark distance $LT$ versus $\varepsilon$ with different values of boost parameter $a$. It can be observed that $LT$ increases monotonically with $\varepsilon$ and reaches a maximum value, $LT_{max}$, at $\varepsilon= \varepsilon_{max}$. This feature suggests that the string is a U-shape type. When $\varepsilon> \varepsilon_{max}$, the $LT$ decrease monotonically with $\varepsilon$, indicating the string configurations solutions no longer satisfy the Nambu-Goto action \cite{Bak:2007fk}. Therefore, we restrict our calculations to the range $\varepsilon_{min}<\varepsilon<\varepsilon_{max}$ ($LT_{min}<LT<LT_{max}$) in the calculations. We also find that the presence of boost parameter suppresses $LT_{max}$, suggesting boost parameter favors the dissociation of quarkonium. Moreover, the dissociation effect is stronger in the transverse cases.

In Fig.~\ref{fig1}(b), we depict the imaginary potential $ImV/(\sqrt{\lambda}T)$ of quarkonium as a function of $LT$ with different boost parameter $a$. The imaginary potential begins at $LT_{min}$ and ends at $LT_{max}$. It is evident that the imaginary potential appears at smaller $LT$ as the boost parameter increases. As noted in Ref. \cite{Finazzo:2014rca}, the suppression will becomes stronger when the imaginary potential occurs at smaller $LT$. Therefore, we can conclude that boost parameter contributes to the melting of quarkonium. Additionally, the impact of boost parameter on the imaginary potential is more significant in the transverse case compared to the parallel case.

The authors of Ref. \cite{Zhang:2023psy} extend the static black hole to a local rotating by the Lorentz transformation and calculate the imaginary potential in this local rotating background.  They find the angular momentum enhances the dissociation. Although the rotating backgrounds considered are different, our results coincide with the results of Ref. \cite{Zhang:2023psy}.

\begin{figure}[H]
    \centering
      \setlength{\abovecaptionskip}{0.1cm}
    \includegraphics[width=8.5cm]{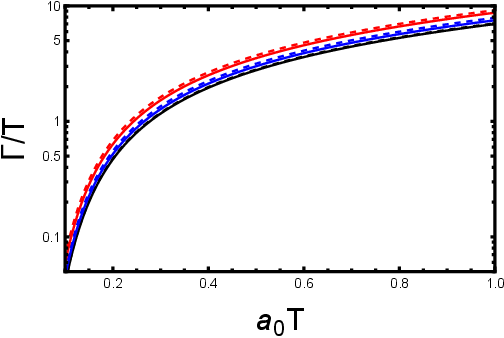}
    \caption{\label{fig2}  Thermal width $\Gamma/T$ versus $a_0 T$ with different boost parameter $a$. The black, blue and red line denote $a = 0.1,\ 0.2,\ 0.3$, respectively. The solid line (dashed line) represents the parallel (transverse) case. }
\end{figure}

Next, we examine how boost parameter affects thermal width. We plot the thermal width $\Gamma/T$ against $a_0 T$ for various boost parameter $a$ in Fig.~\ref{fig2}. One can observe that the boost parameter enhances the thermal width, with a more significant effect observed in the transverse case than in the parallel case. A larger thermal width indicates that the state is unstable and may break down into free quarks, whereas a smaller thermal width suggests a more stable bound state. This phenomenon can be explained by the binding energy of heavy quarkonium. The authors of Ref. \cite{Zhu:2024dwx} investigate the thermodynamics of heavy quarkonium in the spinning black hole background and calculate its binding energy. They discover that the binding energy increases quickly to zero when increasing boost parameter, indicating that heavy quarkonium is more unstable at larger boost parameter. The authors of Ref. \cite{Zhu:2024uwu} examine the spectral function of heavy quarkonium within the same spinning black hole background and find that boost parameter facilitates the dissociation of heavy quarkonium. Our results are consistent with those presented in Refs. \cite{Zhu:2024dwx,Zhu:2024uwu}.

\section{Conclusion and discussion}\label{sec:04}

In this paper, we investigate the imaginary potential and estimate the thermal width from imaginary potential in the spinning black hole background. The imaginary potential could explain the dissociation and decay of the bound state. Also, it can be used to estimate the thermal width related to thermal decay.

We find that the boost parameter suppresses the maximum value of interquark distance $LT$. As boost parameter increases, the imaginary potential appears at smaller $LT$, indicating that boost parameter contributes to the melting of quarkonium. Additionally, we observe that boost parameter enhances the thermal width, indicating that the bound state is more unstable at largerboost parameter. Furthermore, we conclude that boost parameter has a stronger effect on the dissociation of quarkonium when the axis of the quark-antiquark pair $Q\overline{Q}$ is transverse to the direction of boost parameter.

The study of the imaginary potential and the calculation of  thermal width from the imaginary potential under rotation is crucial for understanding their behavior in rotating thermal QGP, which contributes to investigate the quarkonium suppression and thermalization in heavy-ion collisions experiments. Finally, investigating configuration entropy within the spinning black hole may be a meaningful work. We plan to delve into this topic in our subsequent studies.

\section*{Acknowledgments}

%\begin{acknowledgements}
Defu Hou's research is supported in part by the National Key Research and Development Program of China under Contract No. 2022YFA1604900. Additionally, Defu Hou receives partial support from the National Natural Science Foundation of China (NSFC) under Grant No.12435009, and No. 12275104. Zhou-Run Zhu is supported by the startup Foundation projects for Doctors at Zhoukou Normal University, with the project number ZKNUC2023018. Zhou-Run Zhu is also supported by the Natural Science Foundation of Henan Province of China under Grant No. 242300420947.

%\end{acknowledgements}

\end{document}